\documentclass[fleqn,usenatbib]{mnras}

\usepackage{newtxtext,newtxmath}
\usepackage{lastpage}
\usepackage[T1]{fontenc}
\usepackage{ae,aecompl}

\usepackage{graphicx}	% Including figure files
\usepackage{amsmath}	% Advanced maths commands
\usepackage{amssymb}	% Extra maths symbols

\title[sterile neutrino lensing]{The lensing properties of subhaloes in massive elliptical galaxies in sterile neutrino cosmologies}

\author[Giulia Despali et al.]{\parbox{0.9\textwidth}{Giulia Despali$^{1}$\thanks{E-mail:gdespali@mpa-garching.mpg.de}, Mark Lovell$^{2,3}$, Simona Vegetti$^{1}$, Robert A. Crain$^{4}$ and Benjamin D. Oppenheimer$^{5}$ }\\\\
$^{1}$Max Planck Institute for Astrophysics, Karl-Schwarzschild-Strasse 1, 85740 Garching, Germany\\ 
$^{2}$Center for Astrophysics and Cosmology, Science Institute, University of Iceland, Dunhagi 5, 107 Reykjavik, Iceland \\
$^{3}$Institute for Computational Cosmology, Durham University, South Road, Durham DH1 3LE, UK\\
$^{4}$Astrophysics Research Institute, Liverpool John Moores University, IC2, Liverpool Science Park, 146 Brownlow Hill, Liverpool L3 5RF, UK \\
$^{5}$CASA, Department of Astrophysical and Planetary Sciences, University of Colorado, 389 UCB, Boulder, CO 80309, USA \\
}

\date{Accepted XXX. Received YYY; in original form ZZZ}

\pubyear{2019}

\begin{document}
\label{firstpage}
\pagerange{\pageref{firstpage}--\pageref{lastpage}}
\maketitle

\begin{abstract}

We use high-resolution hydrodynamical simulations run with the EAGLE model of galaxy formation to study the differences between the properties of - and subsequently the lensing signal from - subhaloes of massive elliptical galaxies at redshift 0.2, in Cold and Sterile Neutrino (SN) Dark matter models. We focus on the two 7~keV SN models that bracket the range of matter power spectra compatible with resonantly-produced SN as the source of the observed 3.5~keV line. We derive an accurate parametrisation for the subhalo mass function in these two SN models relative to CDM, as well as the subhalo spatial distribution, density profile, and projected number density and the dark matter fraction in subhaloes. We create mock lensing maps from the simulated haloes to study the differences in the lensing signal in the framework of subhalo detection. We find that  subhalo convergence is well described by a log-normal distribution and that signal of subhaloes in the power spectrum is lower in SN models with respect to CDM, at a level of 10 to 80 per cent, depending on the scale. However, the scatter between different projections is large and might make the use of power-spectrum studies on the typical scales of current lensing images very difficult. Moreover, in the framework of individual detections through gravitational imaging a sample of $\simeq$30 lenses with an average sensitivity of $M_\rmn{sub}=5\times 10^{7}\rmn{\rmn{M_{\odot}}}$ would be required to discriminate between CDM and the considered sterile neutrino models.
\end{abstract}

\begin{keywords}
keyword1 -- keyword2 -- keyword3
\end{keywords}

%%%%%%%%%%%%%%%%%%%%%%%%%%%%%%%%%%%%%%%%%%%%%%%%%%

%%%%%%%%%%%%%%%%% BODY OF PAPER %%%%%%%%%%%%%%%%%%

\section{Introduction}

Understanding and unveiling the nature of dark matter is one of the most long-standing challenges in modern astrophysics. According to the standard $\mathrm{\Lambda}$ Cold Dark Matter ($\mathrm{\Lambda}$CDM) model, dark matter 
constitutes the vast majority of the matter content in the Universe and, together with dark energy, accounts for 95 per cent of the total energy budget \citep{planck1_14}. This model has been successful in explaining many aspects of structure formation and evolution, as well as in reproducing the density fluctuations in the early Universe with great accuracy \citep[e.g.][]{wmap9,planck1_14}. However, dark matter models are still untested at the small non-linear scales, due to numerical and observational limitations: a number of unsolved discrepancies exist between N-body simulations based on cold dark matter (CDM) and observations, such as the `too-big-to-fail' and the `core-cusp' problems \citep{klypin99,boylan-kolchin09,bullock17}. 
%Moreover, dark matter models are still untested at the small non-linear scales, due to numerical and observational limitations.  
At the more fundamental level, a dedicated campaign to identify new fundamental particles that fit the requirements for supersymmetric dark matter has not yielded any definitive detections, either indirectly \citep[e.g. the review of][]{gaskins16}, directly \citep[e.g.][]{Aprile18} or via collider searches \citep[][]{aaboud18,sirunyan18}. Thus the combined tension with both astrophysical and particle physics experimental results necessitates the investigation of alternatives. 

Some particle physics models solve the dark matter problem using particles that either evade, or even explain, the problems outlined above. Resonantly produced sterile neutrino dark matter is of particular interest for a series of reasons. It forms part of a well-developed extension of the standard model, called the neutrino minimal standard model ($\nu$MSM), which in addition to providing a plausible dark matter candidate also generates a mechanism to effect neutrino oscillations and baryogenesis \citep{asaka05,laine08,boyarsky09a}. It explains all of these phenomena by introducing just three new, extra particles, which is the minimum required to explain neutrino oscillations and dark matter simultaneously, and could be tested/confirmed by the proposed SHiP experiment \citep{SHiP16}. 

Sterile neutrinos are expected to decay into X-rays at a rate that is accessible to constraint/detection by X-ray observatories, particularly if the sterile neutrino mass is larger than 2~keV \citep{abazajian01a,abazajian01b}. An unexplained X-ray line detected at an energy of 3.55~keV in stacked observations of galaxy clusters \citep{bulbul14}, M31 \citep{boyarsky14a}, the Galactic Centre \citep{boyarsky15,hofmann19} and the Milky Way (MW) halo outskirts \citep{boyarsky18,cappelluti18} counts the decay of a 7.1~keV sterile neutrino among its possible sources. As one of the most promising, although not uncontentious \citep[see][]{anderson15,jeltema16,ruchayskiy16}, indirect dark matter detection signals, it constitutes a viable dark matter candidate, and is especially well suited to further study because the particle physics parameters that determine the X-ray decay signal also set the structure formation properties. 

A set of complementary probes  of the sterile neutrino dark matter cosmology have been derived from the fact that the sterile neutrino particle behaves as warm dark matter (WDM). The term WDM refers to a family of models, including thermal relics and sterile neutrinos, in which there is a cut-off in the linear matter power spectrum at dwarf galaxy scales for viable models; the cut-off scale is at least in part set by the the mass of the WDM particle, and in the specific case of resonantly produced sterile neutrinos can be specified uniquely, notwithstanding systematic uncertainty in the particle physics calculations, by the measured X-ray decay rate and emission energy (see \citealp{abazajian19} for a discussion of additional sterile neutrino models). It has been tested with observations of local dwarf galaxies \citep[e.g.][]{polisensky11,lovell16,schneider17,cherry17}, Lyman-$\alpha$ forest measurements \citep{viel05,viel13,irsic17,garzilli18}, and reionization constraints \citep{bose16b} particularly for low-mass sterile neutrinos; these methods have proven effective at ruling out most resonantly produced sterile neutrino dark matter with a particle mass $<6$~keV.

From the observational point of view, techniques such as the detection of (sub)haloes through gravitational lensing \citep{vegetti09,vegetti09b,vegetti10,vegetti10b,vegetti12,vegetti14,nierenberg14,hezaveh16,hsueh19} or through gaps in the MW stellar streams \citep{erkal16,amorisco16} are very promising for the understanding of the nature of dark matter. Gravitational lensing is sensitive to the whole mass distribution within the lens galaxies and along the line of sight \citep{despali18,vegetti18}: low-mass and non-luminous haloes can be detected via their gravitational effect on the observed lensed images.

The aim of this paper is to study the properties of $z~\sim0.2$ giant elliptical %high redshift 
galaxies and their subhalo populations in sterile neutrino models and make a comparison with CDM, with a particular focus on the implications for strong lensing. For this purpose, we ran high-resolution, hydrodynamical zoom-in simulations of analogues of lens galaxies at $z=0.2$ -- the mean redshift of the SLACS lens sample \citep{bolton06}. This is the first time that hydro simulations have been run of massive ellipticals with sterile neutrinos, while previous works focused on dwarf galaxies and Local Group analogues \citep{lovell17b,bozek19}.

This paper is structured as follows: in Section~\ref{sec_sim} we describe the simulations and the halo selection; in Section~\ref{sec_sub} we present the properties of the subhalo population in CDM and in the two sterile neutrino models, such as the parametrisation of the subhalo mass function, the subhalo density profiles and radial distribution. In Sections~\ref{sec_lensing} and \ref{sec_det} we determine whether these dark matter models can be discriminated using gravitational lensing, by looking at different estimators. In Section~\ref{sec_lensing} we compute the lensing signal of haloes and subhaloes by creating maps of lensing convergence and mock images, with two purposes: comparing the convergence distribution in Section \ref{sec41}, and measuring the subhalo power spectrum in Section \ref{sec42}. Finally, in Section \ref{sec_det} we use the formalism from \citet{li16b} and \citet{despali18} to calculate the expected number of low-mass ($[10^{7}-10^{10}]$~$\rmn{M_{\odot}}$) haloes as a function of mass, lens and source redshift, and observational sensitivity in order to determine if we can discriminate between these dark matter models with current or future lens samples. We summarize our findings, discuss the implications for substructure lensing in detail and draw our conclusions in Section \ref{sec_discussion}.

\section{Simulations} \label{sec_sim}

\begin{table*} 
\begin{center}
\caption{Summary of halo properties in the twelve hydro simulations: ID, halo mass $M_{200c}$ and radius $r_{200c}$, stellar mass of the central galaxy $M_{*}$. In the next four columns we list the total number of subhaloes with {\sc subfind} mass $M>M_\rmn{min}\simeq 3\times 10^{7}\rmn{M_{\odot}} \rmn{h^{-1}}$, those with $M > 10^{8}\rmn{M_{\odot}} \rmn{h^{-1}}$ and the total number of `luminous' satellites, given two different thresholds for stellar mass. For the L8 and L11 runs, we also show in brackets the subhalo abundances in the DMO runs. Finally, we have the stellar effective radius $r_{*,e}$, calculated both from the 3D distribution of stars and averaging the projected distribution over different orientations -- these last two quantities are calculated with the stellar particles belonging to the main galaxy and located within 300 (100) kpc from the halo centre. The masses are in units of $\rmn{M_{\odot}}$, $r_{200c}$ in units of comoving kpc, while the effective radii are expressed in physical kpc, similarly to the observational data in general. ID's 1, 2, 3, and 4 were identified in \citet{oppenheimer16} as Grp008, Grp009, Grp005, and Grp002 in their Table 1. \label{tab_sim1} }
\begin{tabular}{cccccccccc} 
\hline 
ID & $M_{\text{200c}}$ & $r_{\text{200c}}$ & $M_{*}$ & $N_{\text{sub}}$ & $N_{\text{\text{sub}}}(M>10^{8})$ & $N_{\text{sub}}(M_{*}>10^{6})$ & $N_{\text{sub}}(M_{*}>10^{7})$ & $r_{*,e}$(3D) &<$r_{*,e}$(2D)>\\
\hline
& [$\rmn{M_{\odot}}$] & [kpc] & [$\rmn{M_{\odot}}$] & & & & & [kpc] & [kpc] \\
\hline
\multicolumn{10}{c}{CDM} \\
\hline 
1  & 1.06 $10^{13}$ & 401 & 9.45 $10^{10}$ & 2209 & 482 & 84 & 25 &12.76 (8.45) & 9.31 (8.12)\\
2 & 1.05 $10^{13}$ & 401 & 1.00 $10^{11}$ & 2577 & 596 & 112 & 36 & 22.31 (13.31)& 27.11 (13.93)\\
3 & 6.40 $10^{12}$ & 340 &5.07 $10^{10}$ & 1436 & 333 & 61 & 30 & 6.24 (4.64) & 5.80 (4.24)\\
4 & 3.99 $10^{12}$ & 290 &5.48 $10^{10}$ & 829 & 189 & 30 & 16 &6.46 (5.09) & 5.15 (4.09)\\
\hline
\multicolumn{10}{c}{L8} \\
\hline 
1 & 1.03 $10^{13}$ & 397 & 1.03 $10^{11}$ & 815 (818) & 318 (377) & 70 & 20& 10.86 (7.44) & 9.30 (7.00)\\
2 & 1.00 $10^{13}$ & 393 & 1.03 $10^{11}$ & 956 (1070) & 368 (505) & 80 & 31& 12.99 (7.44) & 12.46 (11.35)\\
3 & 6.72 $10^{12}$ & 339 & 7.22 $10^{10}$ & 507 (593) & 200 (246) & 48 & 24& 4.43 (3.71) & 3.64 (2.82)\\
4 & 3.93 $10^{12}$ & 288 & 6.70 $10^{10}$ & 306 (349) & 126 (162)& 27 & 13& 7.03 (5.93) & 6.32 (4.78)\\
\hline
\multicolumn{10}{c}{L11} \\
\hline 
1 & 1.03 $10^{13}$ & 397 & 9.86 $10^{10}$ & 361 (369) & 188 (230) & 63 & 20 & 12.08 (7.83) & 9.30 (7.00)\\
2 & 1.02 $10^{13}$ & 397 & 1.15 $10^{11}$ & 442 (486) & 224 (286) & 73 & 29 & 18.27 (12.85) & 24.68 (12.26)\\
3 & 6.25 $10^{12}$ & 334 & 6.66 $10^{10}$ & 238 (239) & 115 (148) & 42 & 21 & 5.12 (4.17)& 5.38 (3.43)\\
4 & 3.93 $10^{12}$ & 288 & 6.25 $10^{10}$ & 140 (168) & 67 (111) & 25 & 13 & 6.41 (5.19) & 7.38 (4.37)\\
\hline
\end{tabular}
\end{center}
\end{table*}

\begin{figure*}
\includegraphics[width=0.79\textwidth]{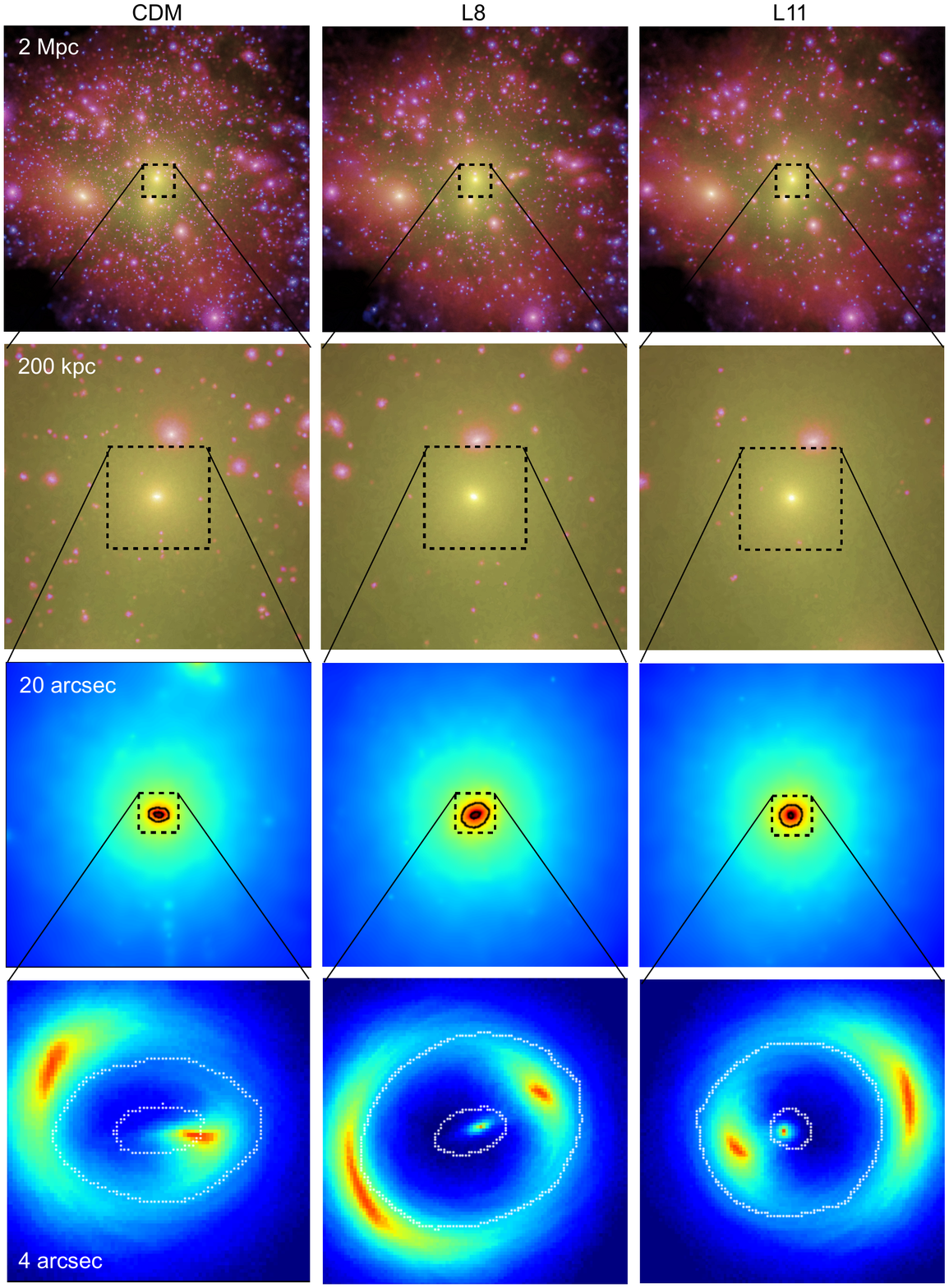}
\caption{Images of one of our simulated haloes (halo 3) at $z=0.2$ that help visualize the scales probed by gravitational lensing. We show the CDM (left-hand column), L8 (middle column) and L11 (right-hand column) models in the full hydro run. In the top row and in the second row we show projections of the dark matter distribution that are 2 Mpc and 200~kpc on a side (physical), respectively: in both cases the images are 2 pMpc deep. Here, image intensity indicates the integrated density along the line of sight, and the colour shows the velocity dispersion. In the third row, the colour scale shows the lensing convergence (in log scale) and the black contours the corresponding critical curves; these projections are 20 arcsec ($\simeq$ 80 pkpc) on a side and were generated by using all of the matter within $r_{200}$. The bottom row shows mock images for the same projection and a simulated source placed at $z=0.6$ (of the size of $\simeq$ 10 pkpc --  or $<3$~per~cent of the virial radius); the colour scale shows the surface brightness or the arc and of the scale usually probed by lensing arcs, while the critical curves are shown again here in white. Even though the halo is seen in the same projection for the three models, the lensing galaxy is not in exactly the same position and does not have exactly the same shape in the three cases: this leads to differences in the projected mass distribution, and thus in the lensed images, when the source is kept in a fixed position. \label{images}}
\end{figure*}

In this section we summarize the halo selection and the details of our simulations.

Our halo selection is derived from the common membership of two parent samples from the Ref-L100-N1504 EAGLE simulation halo catalogue \citep{schaye15}\footnote{The EAGLE catalogues are publicly available at http://icc.dur.ac.uk/Eagle/database.php \citep{mcalpine16}. Galaxies morphologies computed by \citet{thob19} are also available in the database.}. The first parent sample is that of \citet{despali17b}, who made a selection of haloes based on a series of properties that were important for matching observed lens galaxies. These halo properties were the mass (defined as $M_\rmn{200c}$ with respect to the critical density), stellar mass, stellar effective radius and velocity dispersion: observed lens galaxies at $z=0.2$ (such as the SLACS sample) follow a linear relation in the stellar effective radius -- total stellar mass plane, as derived by \citet{auger10b}. Moreover, the selected galaxies were required to have a disc-to-total mass ratio lower than 0.5 in order to select early-type (ETG) morphologies. The second parent sample was assembled by \citet{oppenheimer16} from the same cosmological simulation box. They identified ten group-mass haloes that were suitable for resimulation using zoomed initial conditions, i.e. isolated and of the right mass for their science goals, and subsequently resimulated these at 8-times higher mass resolution than the parent simulation volume. Four of these ten haloes were found to also be a part of the \citet{despali17b} selection. These were selected for resimulation, creating the first sample of massive elliptical galaxies simulated with sterile neutrino cosmologies and baryonic physics. 

The sterile neutrino dark matter cosmology provides for a rich variety of linear matter power spectrum shapes and cut-offs, many of which have already been ruled out by a combination of MW-M31 satellite counts \citep{polisensky11,kennedy14, lovell14,lovell16} and Lyman-$\alpha$ forest studies \citep{viel05,viel13,irsic17}. It is not feasible to run enough simulations to probe in full all of the available parameter space; we therefore take our inspiration from the recently detected 3.55~keV line, one explanation of which is the decay of a sterile neutrino with a mass of 7.1~keV and a neutrino mixing angle  $\sin^2{2\theta}=[2,8]\times10^{-11}$. This range corresponds to a production lepton asymmetry, $L_{6}=[9,11.2]$ \citep{lovell16}. The $L_{6}=9$ model produces a matter power spectrum significantly warmer than that of CDM, and $L_{6}=11.2$ warmer again than $L_{6}=9$.  The coldest matter power spectrum available for a sterile neutrino is attained for $L_{6}=8$, which was found to be consistent with the dark matter decay interpretation of the 3.55~keV line reported by \citet{bulbul14} and \citet{boyarsky14a} but was later ruled out by more stringent limits from \cite{ruchayskiy16}. In order to take account of any possible systematic uncertainty in the calculation of the matter power spectra for these models, we adopt $L_{6}=8$ rather than $L_{6}=9$ as the colder limit for our study, and adopt $L_{6}=11.2$ for the warmer limit. 
We refer to these models hereafter as L11 and L8. In both cases the sterile neutrino mass is 7.0~keV rather than 7.1~keV; we do not expect the results for a 7.1~keV particle would be significantly different from those derived here. 

The simulations are performed in exactly the same way as their CDM counterparts \citep{oppenheimer16}. The galaxy formation model is that of the EAGLE project \citep{schaye15,crain15}, which features cooling, star formation, stellar evolution and feedback, associated with the formation of stars and the growth of black holes. We use the version of the model in which the model parameters are recalibrated for the dark matter particle mass of $1.21\times10^{6}\rmn{M_{\odot}}$, known as RECAL, as do the equilibrium-chemistry simulations in \citet{oppenheimer16}. The code is a heavily modified version of {\sc P-Gadget3} \citep{springel08b} and uses a pressure-entropy formulation of SPH \citep{hopkins13}. Haloes are identified using the friends-of-friends (FoF) algorithm and are subsequently split into subhaloes using the {\sc subfind} halo finder \citep{springel01b}. The cosmological parameters are consistent with the constraints by the \citet{planck1_14}: $\Omega_{0}=0.307$, $\Omega_{\Lambda}=0.693$, $\Omega_\rmn{b}=0.04825$, $h_{0}=0.6777$, $\sigma_8=0.8288$ and $n_\rmn{s}=0.9611$. The only manner in which these simulations differ from their CDM versions is that the initial conditions have been remade with the L11 and L8 matter power spectra, which are shown in fig.~1 of \citet{lovell17b}; like the CDM simulations, our runs were simulated with mass resolution eight times higher than the simulation box from which they were sourced. Moreover, for the sterile neutrino models, we ran both full-hydro and  dark matter-only (DMO) versions for each of the four volumes.

Figure~\ref{images} shows images of one of our haloes at $z=0.2$, which present qualitatively the differences between the CDM, L8 and L11 models at different scales. In the top row we show projections of the dark matter distribution that are 2 physical Mpc on a side of the CDM (left-hand column), L8 (middle column) and L11 (right-hand column) halo 3 runs. The second row shows instead the central 200 pkpc. In the third row the colour scale represents the lensing convergence -- defined as the projected mass density distribution normalized to the critical density, see Section \ref{sec_lensing} -- in a box of 80 pkpc ($\simeq$20 arcsec) on a side, created by using all of the matter within $r_\rmn{200c}$, i.e. baryonic and dark matter combined. Finally, the last row shows mock lensing images, created by placing a source at $z=0.6$: lensed images appear in the very central part, on a scale of the order of $\simeq$2 arcsec/$\simeq$ 10 kpc --  or $<3$~per~cent of the virial radius. 

The lowest mass subhalo -- i.e. the bound {\sc subfind} mass, $M$ -- that is resolved in our simulations is  $M\simeq 3\times 10^{7}\rmn{M_{\odot}}\rmn{h^{-1}}$ (20 particles, which is the default minimum for {\sc subfind}), but a robust identification is possible only at $M\geq 10^{8}\rmn{M_{\odot}}\rmn{h^{-1}}$. At this scale, WDM simulations suffer from spurious fragmentation of filaments, due to the presence of a resolved wavenumber cut-off in the initial power spectrum. These fragments may then be identified as subhaloes by the subhalo finder, artificially increasing the abundance of the low mass objects. This is a purely numerical problem, which is not solved sufficiently by increasing the spatial resolution, but which might be alleviated by future N-body codes that use phase space smoothing techniques \citep{angulo14} or instead apply an adaptive softening \citep{hobbs16}. At present, the only solution is to exclude them from the analysis using empirical criteria: in this work we follow the method developed by \citet{lovell14}, which enables us to eliminate spurious subhaloes based on their peak mass and their Lagrangian shape in the initial conditions, with specific thresholds for each sterile neutrino model (see their work for details on how these are derived). In what follows, we always make use of  subhalo catalogues from which the spurious subhaloes have been removed. In all cases, the total mass fraction in spurious subhaloes with respect to $M_{200c}$ is less than 0.0002 and thus does not affect the results.

\subsection{Properties of the host haloes}

As can be seen from Table \ref{tab_sim1}, the properties of the main haloes and also of the central galaxies are very similar in all three models (CDM, L8 and L11). The values of $M_{200c}$ and the total stellar mass differ by only a few per~cent, in agreement with the results from \citet{lovell17b}. The main difference between these dark matter models is indeed the number of subhaloes, as we will discuss in the next section. However, we caution that any change in the galaxy formation model will change both the overall matter distribution \citep{springel18} and thus the disruption rate of satellites \citep{richings19}, which has to be borne in mind when interpreting our results. Recently, \citet{ludlow19} showed that small galaxies and the inner parts of galaxies in hydrodynamical simulations might be affected by numerical artefacts due to the energy transfer from dark matter to stars when two species of particles with unequal mass are present (equipartition), as is the case in our runs. This has an effect on galaxy sizes and implies that the softening is not a reliable estimate of the minimum reliable size. In principle, this kind of numerical effect could have an effect on the estimate of the lensing signal from haloes and subhaloes, which originates in the central parts of galaxies, as well as on the estimate of density profiles and in general halo properties close to the centre. The softening length of our simulations is $\epsilon=0.350$ physical kpc and thus an effective resolution limit is  $2.8\times \epsilon=0.98$ pkpc. We mark these values in all the relevant figures in order to delimit the parameter space where the results have to be interpreted with particular care.

In the main body of the paper we focus on the comparison between CDM and sterile neutrino models, in particular on the relative numbers of subhaloes and thus their relative impact on the number of satellites: any numerical artefacts due to the equipartition effect likely affect all three sets of simulations to a similar degree, and thus we expect that our results on the relative contribution of subhaloes are not strongly affected.

\citet{sampath19} studied the effect of different feedback mechanisms on the lensing signal of ETGs. They used a suite of $\Lambda$CDM hydrodynamical simulations, run with different variations of the EAGLE galaxy formation model, to measure the density slope at the Einstein radius and the projected mass-density relation, and how these depend on different implementations of the stellar and AGN feedback. They argue that, through a detailed comparison with observed samples, it is possible to use gravitational lensing to constrain feedback models. Here we can complement their findings, identifying that, since the main halo properties are not systematically affected by the change in the dark matter model (see Table \ref{tab_sim1}), the lensing properties of host haloes cannot necessarily be used to distinguish CDM from sterile neutrino models (at fixed baryonic physics model).

\begin{figure}
    \centering
    \includegraphics[width=0.95\hsize]{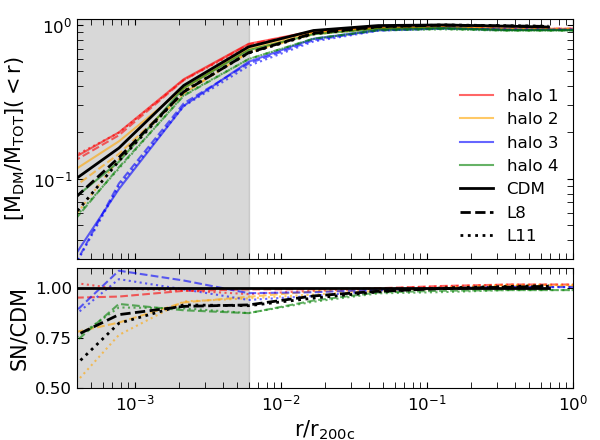}
    \caption{Dark matter fraction of the four simulated haloes as a function of distance from the centre, expressed as the fraction of the virial radius $r_\rmn{200c}$. The coloured solid (dashed and dotted) lines show the CDM (L8 and L11) scenarios for individual haloes, with black lines showing the mean relations given that there is a significant variation among the haloes. The lower panel shows the ratio between the sterile neutrino mean relations and CDM. The grey shaded region marks the length scale below the effective spatial resolution of the simulation ($2.8\times \epsilon$), where results have to be interpreted with care.}
    \label{mass_frac}
\end{figure}

Another property that we can measure is the dark matter fraction of the haloes -- which has been measured in hydrodynamical CDM simulations \citep{schaller15,lovell18} -- in order to see how it differs in sterile neutrino models. Figure \ref{mass_frac} shows the dark matter fraction, as a function of distance from the centre, for the four simulated haloes (lines of different colour). For each halo, the dark matter mass fraction is hardly distinguishable in the three models (solid, dashed and dotted lines of the same colour) for $r>0.05r_\rmn{200c}$. A small suppression ($\simeq$ 20 per cent), indicating a smaller dark matter fraction at the very centre of the haloes in both sterile neutrino models, is apparent in the mean distribution; however, we caution that we do not have enough statistics to provide reliable estimates, given the halo-to-halo fluctuation. We can conclude that the overall mass distribution of the four haloes is similar in CDM and sterile neutrino models.  This is consistent with the fact that their properties, such as total mass, stellar mass and radius, agree very well (see Table \ref{tab_sim1}). Moreover, it supports the fact that the mass accretion is very similar in the three models, but in CDM a higher percentage of the accreted mass is in the form of clumpy structures. As shown by \citet{schaller15}, the dark matter fraction increases with halo mass in the central region of the halo, which explains the differences between the four haloes. 

\section{Subhalo properties} \label{sec_sub}

In this section we analyse four dark matter-dependent subhalo population properties that could play a role in setting the gravitational lensing signal of subhaloes: the mass functions, the spatial distribution, the density profiles and the dark matter fraction in subhaloes.

\subsection{Mass function} \label{sec_mf}

\begin{figure}
\includegraphics[width=\hsize]{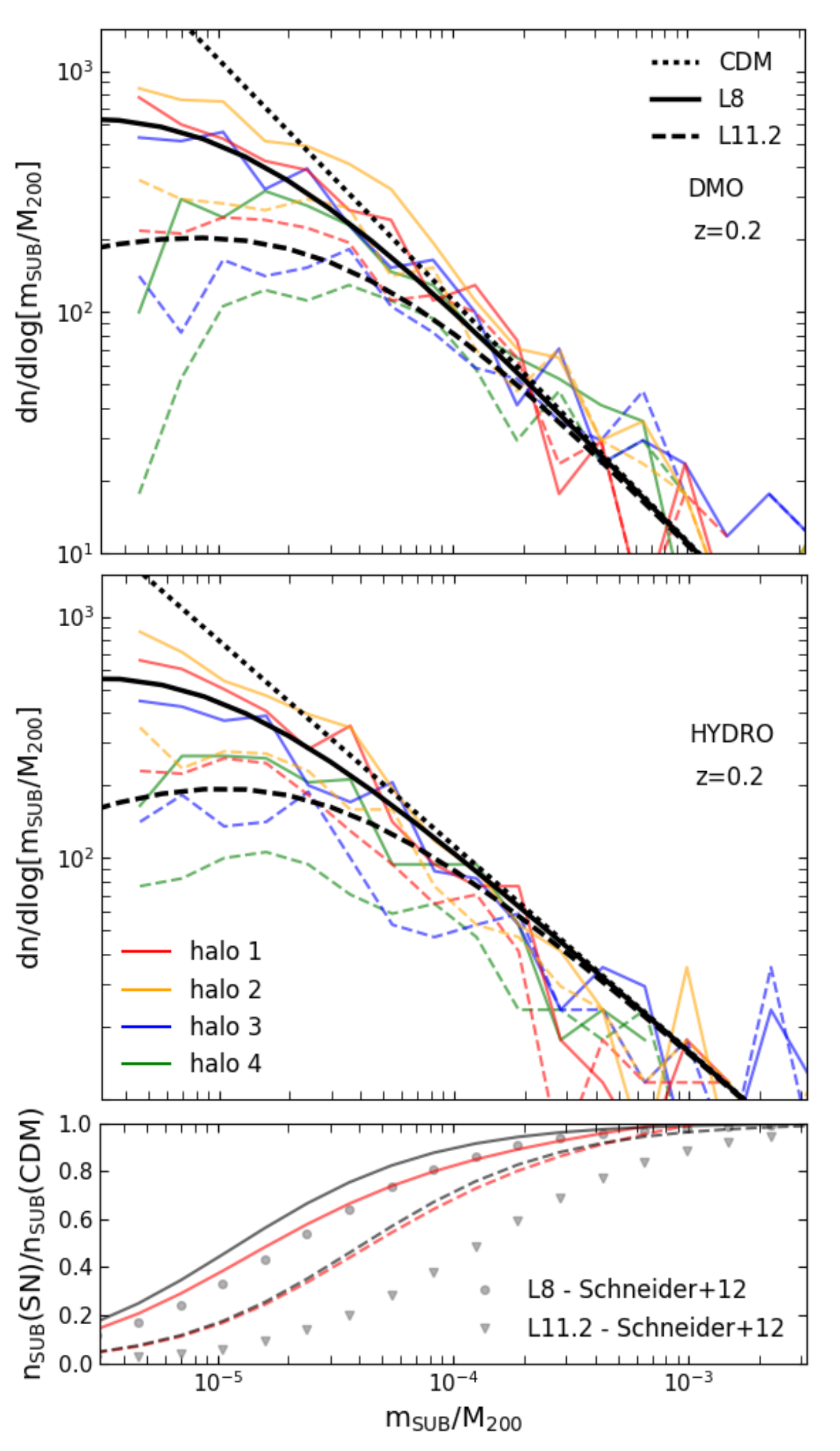}
\caption{Subhalo mass function at $z=0.2$. \emph{Upper and middle panels}: the mass function of each halo is represented by a different colour, with solid (dashed) lines standing for the L8 (L11) subhalo mass function. The black solid and dashed lines show the best fit to the mean subhalo mass function for the two cases, while the black dotted line shows the CDM mass function derived in \citet{despali17b}. The upper panel shows the results from the dark matter-only run, and the middle panel shows results from the full-hydro counterpart. \emph{Lower panel}: we show the ratio in subhalo counts between the  sterile neutrino models and the CDM scenario. Here the solid (dashed) lines correspond to L8 (L11), while the grey and red colours correspond respectively to the dark matter only and hydro runs. The grey circles and triangles show the subhalo mass function predicted according to the mass function fit by \citet{schneider12}, given the values of $M_\rmn{hm}$ of the two sterile neutrino models.  \label{submf}}
\end{figure}

The formation of low-mass ($<10^{10}$~$\rmn{M}_{\odot}$) dark matter haloes is suppressed due to the cut-off in the initial matter power-spectrum, and the subhalo number density may experience an additional reduction due to the interaction with the host halo, which could feasibly vary between models due to changes in the mass--concentration relation.
Nevertheless, previous works have assumed that (to the first order approximation) the degree of suppression is the same or similar for isolated haloes and subhaloes. In particular,  \citet{schneider12} parametrised the ratio between thermal relic WDM and CDM isolated halo number density as a function of mass as 
\begin{equation}
\frac{n_\rmn{WDM}}{n_\rmn{CDM}} = \left(1+ \frac{M_\rmn{hm}}{M_\rmn{sub}} \right)^{\beta},
\label{massf_eq_1}
\end{equation}
where $M_\rmn{hm}$ is the half-mode mass scale  as defined by the ratio of the WDM linear matter power spectrum relative to CDM, and $\beta$=-1.16. For the case of MW subhaloes, \citet{lovell14} found a better fit  with $\beta=-1.3$ or with a slightly different parametrisation:
\begin{equation}
\frac{n_\rmn{WDM}}{n_\rmn{CDM}} = \left(1+ \gamma\frac{M_\rmn{hm}}{M_\rmn{sub}} \right)^{\beta},
\label{massf_eq_2}
\end{equation}
where $\beta=-0.99$ and $\gamma=2.7$. It is important to point out that most studies that try to distinguish between different dark matter models using the subhalo counts \citep[e.g.][]{birrer16,li16b,despali18,vegetti18,hsueh19} use the parametrisation from \citet{schneider12} for any kind of WDM model. This choice is motivated by the fact that  recent works focused on thermal relic WDM candidates \citep[e.g.][]{viel05,lovell12,lovell14,bose16b}, for which this parametrisation was originally developed. As discussed in Section \ref{sec_sim}, sterile neutrino models are intrinsically more complex, given that they are characterized by the sterile neutrino mass $m_{s}$ and the lepton asymmetry $L_{6}$. The combination of the two determines %the model half-mode mass 
$M_\rmn{hm}$, while for thermal relic candidates this is determined by the particle mass alone. $M_\rmn{hm}$ has a value of $1.28\times10^{8}\rmn{M}_{\odot}$ ($8.25\times10^{8}\rmn{M}_{\odot}$) for LA8 (LA11). For the \citet{viel05} fit to the thermal relic power spectrum these $M_\rmn{hm}$ correspond to thermal relic particle masses of 4.3~keV and 2.5~keV for LA8 and LA11 respectively.

 Note that for the purpose of the fitting formula in \citet{lovell14}, we should multiply the `true' values of $M_\rmn{hm}$ by the $\gamma$ parameter to derive an `effective' half-mode mass; in this case our LA8 and LA11 models correspond to `effective' thermal relic masses of 6.0 and 4.2~keV respectively. However, it is also the case that the sterile neutrino matter power spectra show a greater variety of cut-off gradients than thermal relic particles, and therefore even knowing $M_\rmn{hm}$ is not sufficient to define the subhalo mass function uniquely. Moreover, we stress that different particle physics codes predict different sterile neutrino momentum distribution functions and consequently different matter power spectra. We performed an approximate re-calculation of our matter power spectra using the public particle physics code published by \citet{venumadhav16} together with the {\sc class} Boltzmann solver \citep{blas11}, which returned $M_\rmn{hm}$ for LA8 (LA11) that were $\sim6$ ($\sim3$) times higher than those found for our spectra; we defer a detailed comparison to future work.

We fit the ratio between the subhalo mass functions in our simulations with the functional form from Equation~(\ref{massf_eq_2}). Figure \ref{submf} shows the subhalo mass function from our DMO (top panel) and hydro (middle panel) runs. The three black curves show the best fit to the subhalo mass function for each dark matter model. In order to obtain a good fit, $\gamma$ has to change for each sterile neutrino model: the best fit values are summarized in Table \ref{tab_par}. The CDM mass function with the same normalization and with slope ($\alpha=0.9$ for the dark matter-only run and $\alpha=0.85$ for the hydro run, as derived in \citealp{despali17b} for the EAGLE hydro simulation) is shown by the dotted line.
The lower panel of Figure \ref{submf} shows the ratio with respect to the CDM DMO scenario. In the same panel, the grey circles and triangles show the subhalo mass function predicted according to the mass function fit by \mbox{\citet{schneider12}}, which underestimates the number of subhaloes for our models by up to 30~per~cent \footnote{The difference between our sterile neutrino models and the fit from \citet{schneider12} might reside in the combination of different aspects: $(i)$ the shallower power-spectrum cut-off in sterile neutrino models with respect to thermal relic WDM; $(ii)$ differences between the halo and the subhalo mass functions; $(iii)$ numerical effects due to different ways of removing the spurious subhaloes (of secondary importance).}.
The same parametrisation works well for the subhalo population at other redshifts (not shown here), once one takes into account the evolution of the normalisation of the mass function with time \citep{giocoli08a}.

The resolution of our simulations allows us to probe the subhalo mass function reliably only down to $M\simeq  10^{8}\rmn{M_{\odot}}\rmn{h^{-1}}$, where it is suppressed by $\simeq$ 50 (80) per cent in the L8 (L11) model relative to the CDM subhalo mass function but is still mostly flat. However, extending the functional forms from Figure \ref{submf} (solid and dashed black lines) to lower masses would result in a sharp decrease at $M<  10^{7}\rmn{M_{\odot}}\rmn{h^{-1}}$. It remains to be shown definitively whether this drop-off rate describes the sterile neutrino cosmology accurately, and confirmation will require still higher resolution simulations.

Finally, it is important to remember that the number of luminous satellites is another important probe, since any viable dark matter model must be able to reproduce their observed abundance. As shown in \citet{lovell16}, some sterile neutrino models are already ruled out by satellite counts in MW-like haloes. Given that the total mass of our haloes is larger than for the MW, it is worth investigating the abundance of satellites when they are separated into luminous and dark sub-populations. In Table \ref{tab_sim1} we list the total number of subhaloes that have stellar masses $M_{*}>10^{6}\rmn{M_{\odot}}\rmn{h^{-1}}$ (lowest resolved stellar mass) or $M_{*}>10^{7}\rmn{M_{\odot}}\rmn{h^{-1}}$: it is easy to see that the total number are slightly lower in the WDM models, but comparable to CDM\footnote{We use these stellar masses only in order to provide a rough estimate of dark vs. luminous satellites; as these masses are at the resolution limit of our simulations, the exact values should not be used for interpreting our results any further.}. This again supports the importance of strong lensing as a method to discriminate between different dark matter models, given its ability to detect the total mass and thus the majority of WDM subhaloes. 

\begin{table} 
\caption{Best-fit mass function parameters from this and previous works \citep{schneider12,lovell14}, following the parametrisation from Eq. \ref{massf_eq_2}.   \label{tab_par}}
\begin{tabular}{ccccc} 
\hline 
\multicolumn{5}{c}{Mass function parameters} \\
model & source & $\gamma$ & $\beta$ & $M_\rmn{hm}[\rmn{M_{\odot}}]$ \\
\hline 
L8-DMO & this work & 0.53 & -1.3 & $1.28\times10^{8}$\\
L11-DMO & this work & 0.27 & -1.3 & $8.25\times10^{8}$\\
L8-HYDRO & this work & 0.35 & -1.3 & $1.28\times10^{8}$\\
L11-HYDRO & this work & 0.18 &  -1.3 & $8.25\times10^{8}$\\
WDM(th. rel.) & Schn+12 & 1 & -1.16 & - \\
WDM(th. rel.) & Lov+14  & 1 &-1.3 & - \\
WDM(th. rel.) & Lov+14(sub) & 2.7 & -0.99 & - \\

\hline
\end{tabular}
\end{table}

\subsection{Radial distribution}\label{sec_rad}

\begin{figure*}
\includegraphics[width=0.47\hsize]{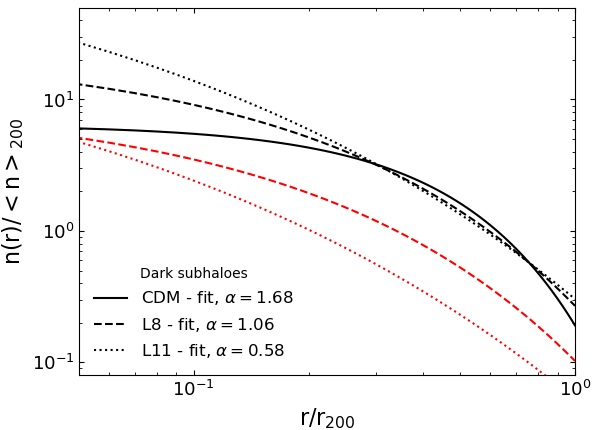}
\includegraphics[width=0.48\hsize]{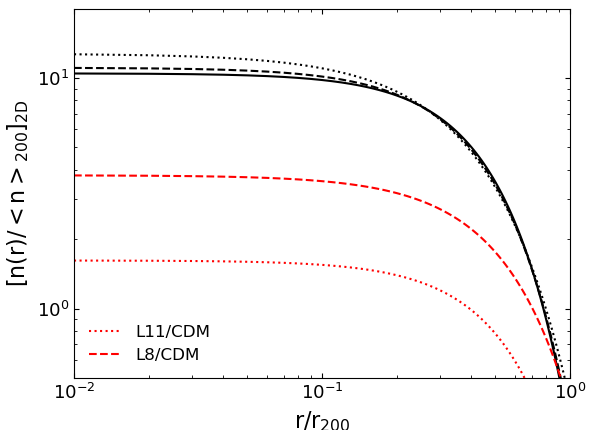}

\caption{Mean radial number density of subhaloes in the hydro runs, normalized to the mean number density within $r_{200}$, for all subhaloes with mass $M>2\times 10^{7}\rmn{M_{\odot}}\rmn{h^{-1}}$. The left-hand panel shows the three-dimensional radial number density, while the right-hand panel presents the projected number density, averaged over three projections for each halo (black lines). Different line styles show the mean number for CDM (solid), L8 (dashed) and L11 (dotted), while the best Einasto profile fit $\alpha$ is given in the figure legend.
The red dashed and dotted lines show the median number density in sterile neutrino models, when normalized to the mean number density in CDM, instead of their own. 
\label{sub_dist}}
\end{figure*}

As shown by \citet{bose17}, the subhalo radial distribution might be different in WDM models with respect to CDM. In particular, even though less numerous, subhaloes in WDM scenarios can be more centrally concentrated than in CDM. In the left-hand panel of Figure \ref{sub_dist} we plot the mean radial number density of subhaloes as a function of radius for our three scenarios, rescaled to the mean number density within $r_{200c}$ in each model. The solid, dashed and dotted curves show the best-fit Einasto profiles to the mean number densities. As expected, the subhalo population is more centrally concentrated than in CDM.  For all three cases $r_{-2}\simeq 0.54$~$r_{200c}$, while the logarithmic slope is different for each model and decreases for warmer models. If the sterile neutrino models are normalized by the CDM mean values (red dashed and dotted lines in both panels), it is apparent how the subhalo population is suppressed in warmer models. 

However, since gravitational lensing is sensitive to projected quantities, we also need to test to what degree this difference in the three dimensional radial distribution is preserved in projection. 
In the right-hand panel of Figure \ref{sub_dist}, we show the projected number density of subhaloes as a function of radius, normalized by its value at $r_{200c}$, by averaging over three different projections for each halo. We see that, even though the normalized number density is still higher for the sterile neutrino models, the mean distributions flatten close to the centre (i.e. inner most $\simeq$30 kpc) and the central number densities are of the same order of magnitude for the three models.

\subsection{Density profiles}\label{sec_prof}

Another recognised effect of WDM is to produce shallower density profiles for the low-mass subhaloes, in the same mass range in which the number of structures is already suppressed \citep[e.g.][]{lovell12,ludlow16}. It has been shown \citep{springel08a} that subhaloes in CDM can be well described both by NFW \citep{navarro96} and Einasto \citep{einasto65} profiles. Here, we use the second since it provides a more flexible parametrisation which might be needed in warmer models.

Thus, we select the subhaloes within $r_{200c}$ and fit their profiles with Einasto profiles, characterized by a power-law logarithmic slope:
\begin{equation}
\eta(r) = \frac{\mathrm{d}\ln\rho}{\mathrm{d}\ln r}(r)\propto r^{1/\alpha}.
\end{equation}
In this model, the integrated density profile is commonly written as
\begin{equation}
\rho(r) = \rho_{-2}\exp\left\{-2\alpha\left[\left(\frac{r}{r_{-2}}\right)^{1/\alpha} -1\right]\right\}.
\end{equation}
We perform the fit by optimizing for the three parameters ($\rho_{-2},r_{-2},\alpha$) for each subhalo, where the first two are the density and the radius at which $\rho(r)\propto r^{-2}$ and $\alpha$ defines the steepness of the power-law. Instead of fitting the individual subhalo profiles -- which could be more prone to uncertainties -- we stack subhaloes of the same mass and we fit the median profile. 

$\rho_{-2}$ is 20 (40) per cent lower in the L8 (L11) model, while $r_{-2}$ is roughly 10 (20) per cent larger at fixed $M_\rmn{sub}$. We estimate the subhalo concentration $c=r_\rmn{max}/r_{-2}$, i.e. the ratio between the subhalo size (or the distance of the farthest particle) and $r_{-2}$. $r_\rmn{max}$ is on average the same in the three models, as a function of mass and, at fixed mass, of the same order of magnitude of the virial radius.

\begin{figure}
    \centering
    \includegraphics[width=\hsize]{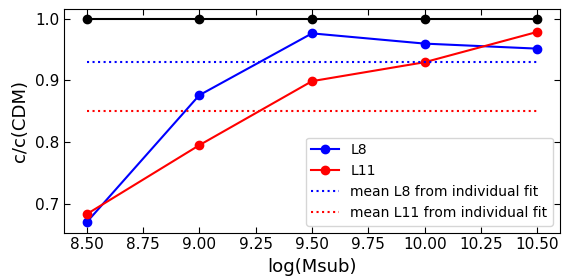}
    \caption{Subhalo concentration decrease  of sterile neutrino subhaloes with respect to CDM, as a function of subhalo mass. The horizontal dotted lines show the mean suppression calculated with the fit to each subhalo.}
    \label{prof_fit2}
\end{figure}

In Figure \ref{prof_fit2}, we show the ratio of the concentration--mass relations of sterile neutrino models relative to CDM. The blue and red lines show the ratio for the L8 and L11 scenarios for each mass bin, while the mean suppression obtained by fitting individual profiles is shown by the dotted lines. 

\begin{figure}
    \centering
    \includegraphics[width=\hsize]{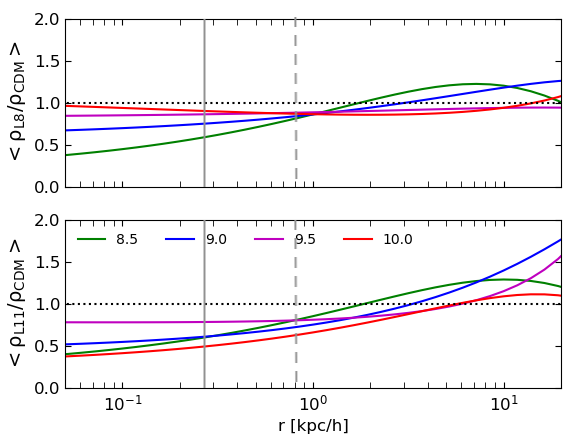}
    \caption{Median density profile ratio with respect to CDM. We use the median values of the Einasto profile parameters to compare the density profiles in the different models, for four subhalo mass bins (different colours). The upper (lower) shows the ratio between the density profile in the L8 (L11) model. The solid and dashed vertical lines mark the softening length of the simulation $\epsilon$ and the effective resolution limit ($2.8\times \epsilon$).}
    \label{prof_fit3}
\end{figure}

We then generate the density profile that would be predicted from the medians of these distributions for four bins in subhalo mass in which we have enough statistics. In  Figure \ref{prof_fit3} we plot the ratio between the predicted density profile in the L8 (upper panel) and L11 (lower panel) models and CDM. The median parameters from the sterile neutrino models yield central densities ($r<1$~kpc) that are systematically lower than the CDM ones, to 30 and 50 per cent at the very centre, consistently for all mass bins. 

\subsection{Dark matter fraction in subhaloes}\label{sec_dmfrac}

Many previous works studied the dark matter fraction in subhaloes \citep{gao04b,springel08b,xud15,despali17b}. 
In WDM models, the halo concentrations are lower \citep[][-- and as in the previous section]{maccio13,ludlow16} and the number of dark subhaloes is suppressed, which can lead to a lower fraction of the total halo mass being located in dark matter subhaloes.

\begin{figure}
    \centering
    \includegraphics[width=\hsize]{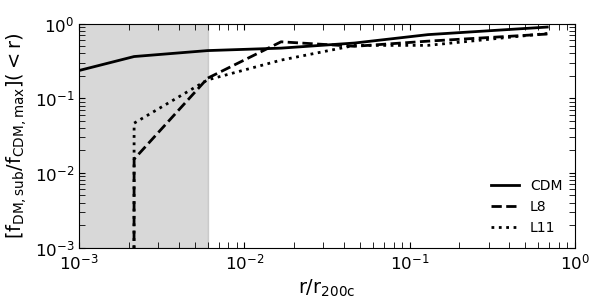}
    \caption{Dark matter fraction in subhaloes of the four simulated haloes, as a function of distance from the centre, expressed as the fraction of the virial radius $r_\rmn{200c}$. The grey shaded region marks the length scale below the effective spatial resolution of the simulation ($2.8\times \epsilon$), where results have to be interpreted with care.}
    \label{mass_frac2}
\end{figure}

In Section \ref{sec_sim} we discussed the total dark matter fraction of the host haloes; we now calculate the dark matter fraction in subhaloes. Its value depends not only on the distance from the centre, but also on halo mass \mbox{\citep{gao04,springel08b}}. For this reason, and since our haloes span almost one order of magnitude in mass, we show the ratio of the sterile neutrino and CDM fractions instead of the absolute value, and present the results in Figure~\ref{mass_frac2}.  As expected from the suppressed mass function, in sterile neutrino models the fraction is lower than in CDM.
The suppression rises from an order of magnitude in the centre to roughly 20 per cent at the virial radius. As shown in previous works \citep[e.g.][]{springel08b,despali17b}, the number density of subhaloes increases towards the centre, in a way that is well represented by an Einasto profile \citep{ludlow09}, more slowly than the total dark matter density and thus most subhaloes are found in the outer parts of the halo. Given that these are present in similar numbers in all the three runs, the dark matter fraction in subhaloes is similar in the outskirts of the halo. On the other hand, the low-mass subhaloes, which can be found closer to the centre, are more abundant in CDM than in sterile neutrino models, and thus the fraction in subhaloes is suppressed towards the centre, even though the number densities of subhaloes centrally is similar between CDM and WDM.

\begin{figure}
    \centering
    \includegraphics[width=\hsize]{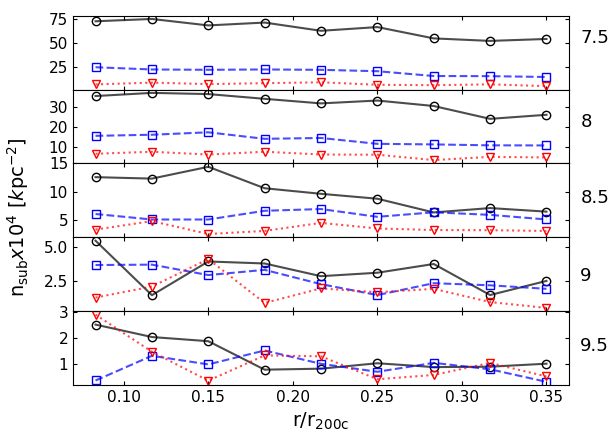}
    \caption{Average projected number of subhaloes per unit area as a function of distance from the centre. We show the results for five mass bins in subhalo mass -- $\log(M/\rmn{M_{\odot}})$ = $(7.5,8,8.5,9,9.5)\pm0.25$ going from top to bottom -- while we average over the whole halo sample and different projections of each halo; the solid black (dashed blue and dotted red) lines, show the results for the CDM (L8 and L11) run.}
    \label{mass_frac3}
\end{figure}

The two sterile neutrino models give similar predictions and cannot be distinguished with only four simulated systems. However, \citet{lovell14} also reported a similar suppression for the slightly different thermal relic WDM model in the MW halo context. 
Finally, we measure the projected subhalo mass fraction and the projected number of subhaloes as a function of the distance from the centre, thus bringing together the findings from previous subsections. In this case, we use different projections for each halo in order to gain more statistics. \citet{xud15} and \citet{despali17b} have shown that the projected number density of subhaloes as a function of radius is constant at small distances from the centre ($r<0.3r_\rmn{200c}$), while it slowly decreases outside. Moreover, the projected number density in CDM is inversely proportional to the subhalo mass, i.e. low-mass subhaloes are more abundant than high mass haloes. This behaviour is obviously modified in sterile neutrino models.
Figure \ref{mass_frac3} shows the projected number density of subhaloes per kpc$^{-2}$ as a function of distance from the centre, focusing on a specific bin in subhalo mass in each panel. The projected density of sterile neutrino models (dashed and dotted lines for L8 and L11, respectively) are of the same order of the CDM ones (solid lines) for the two higher mass bins ($M=10^{9-9.5}\rmn{M_{\odot}}$), while they are 50 per cent lower for $M=10^{8-8.5}\rmn{M_{\odot}}$ and 70 per cent lower for the lowest bass bin available in our simulations ($M=10^{7.5}\rmn{M_{\odot}}$).

\section{Lensing properties of subhaloes} \label{sec_lensing}

In this section, we analyse the lensing properties of the subhalo population in detail. We look at the lensing signal directly by creating mock images and lensing convergence maps, and subsequently measuring the distribution of subhalo convergence and the corresponding power spectrum.

We use the lensing code {\sc GLAMER} \citep{metcalf14,petkova14} to run ray-tracing simulations and generate lensing maps of the simulated haloes. 

For any extended distribution of matter, the lensing effect depends on the projected mass distribution and so is characterized by the effective lensing potential, which is in turn obtained by projecting the three-dimensional potential $\Phi$ on the lens plane:
\begin{equation}
\Psi(\theta) = \frac{D_\rmn{LS}}{D_\rmn{L}D_\rmn{S}}\frac{2}{c^{2}}\int\Phi(D_\rmn{L},\theta,z) \mathrm{d}z,
\end{equation}

\noindent where $D_{\text{L}}$, $D_{\text{S}}$ and $D_{\text{LS}}$ are, respectively, the angular diameter distance of the observer to the lens, the observer to the source, and between the lens and the source; $c$ is the speed of light. The gradient of the lensing potential gives the scaled deflection angle, while the Laplacian gives the convergence $\kappa$:
\begin{equation}
\Delta_{x}\Psi(x)=2\kappa(x).
\end{equation}
The lensing convergence is defined as a dimensionless surface density and so effectively corresponds to a scaled projected mass density, characterising the lens system. It can be written as
\begin{equation}
\kappa(x)= \frac{\Sigma(x)}{\Sigma_\rmn{crit}}, \qquad \rmn{with} \qquad \Sigma_\rmn{crit}= \frac{c^{2}}{4\pi G} \frac{D_\rmn{S}}{D_\rmn{LS}D_\rmn{L}},
\end{equation}
where $\Sigma_\rmn{crit}$ is the critical surface density. The value of the lensing convergence in practice determines by how much the background sources appear magnified on the lens plane. For non-spherical mass distributions, the lensed images of the source are also stretched and distorted along privileged directions by the shear, $\gamma$. 
Finally, the critical curves identify the location on the lens plane in which the magnification is ideally infinite (see the white contours in Figure \ref{images} for an example).

{\sc GLAMER} calculates the deflection angles, shear and convergence by a tree algorithm, representing each simulation particle with a B-spline in three dimensions as is commonly done in smooth particle hydrodynamics (SPH) simulations.  The size of the particles is set to the distance to the $N$th nearest neighbour where $N_\rmn{smooth}$ can be adjusted; in this work $N_\rmn{smooth}=64$.  This smoothing scheme provides higher resolution where the particles are dense and higher resolution is therefore justified, and lower resolution where the particles are sparse and shot noise would otherwise be a problem. We set the lens and source redshift respectively to $z_\rmn{l}=0.2$ and $z_\rmn{s}=0.6$. For the sterile neutrino runs, we exclude the particles belonging to spurious subhaloes. 

We ran the ray-tracing code through 200 random projections for each halo, in order to obtain a good statistical sample. We look at the lensing convergence generated by subhaloes ($\kappa_\rmn{sub}$). We exclude those subhaloes that have a significant stellar component ($M_{*}>10^{7}\rmn{M_{\odot}}\rmn{h^{-1}}$ - see Section \ref{sec_mf}), as we are interested in the dark subhaloes that cause perturbations to the lensing signal but cannot be detected through their light. This is in practice equivalent to removing the same number of massive subhaloes from the maps and has only minor consequences for the distribution of convergence, since it only influences the high-$\kappa_\rmn{sub}$ tail, but not the shape or mean of the distribution. 

In particular, we focus on maps of 100 arcsec on a side (thus containing the whole radius of the haloes in projection) and of 10 arcsec on a side, slightly larger than the typical lensing images. The two sets of maps have the same resolution of 0.1 arcsec (similar to that of HST data) and the same depth along the line of sight (i.e. $2\times r_\rmn{200c}$). In this way we  want to see how much of the `true' signal from subhaloes can be extracted from individual lensing images, which enclose a region that is much smaller than the virial radius on the plane of the sky (see Figure \ref{images} for an example of the relevant scales). The top panels of Figure \ref{conv} show an example of CDM total convergence maps of the two considered fields-of-view. In the left-hand panel, the subhaloes can be identified by eye, while the central region on the right is dominated by the main galaxy and only one extra clump is clearly visible, proving once more the challenge of finding subhaloes with lensing. 

We proceed by analysing the distribution of the subhalo lensing convergence and its power spectrum. Previous works have tried to describe the effect of subhaloes on lensed arcs as the presence of residuals (i.e. the difference between the observed lensed images and those predicted by the best lens model) in the form of a Gaussian random field \citep{bayer18,saikat18}. Moreover, a number of recent works focused on the subhalo power spectrum, measured from the lensing convergence, using analytical (sub)haloes or numerical simulations comparing cold, warm (thermal relic) models \citep{cyr-racine2018,brennan18} or also self-interacting dark-matter models \citep{rivero18}. 

Previous works analysed the dependence of the subhalo power spectrum on both the maximum subhalo mass and the size of the halo sample (and number of projections). While they use analytical density profiles, here we are instead intrinsically limited by the resolution and the number of the simulations. Thus, we can neither characterize fully the variance due to differences between the haloes, nor explore the effect of the substructure population below $M\simeq 10^{8}\rmn{M_{\odot}}\rmn{h^{-1}}$ reliably. Moreover, the sterile neutrino models used in this work are colder than the model used in their work and so the difference in the power spectra is unlikely to be as pronounced. Nevertheless, we can explore the variance due to different projections and -- since the same haloes have been run with different dark matter models -- estimate how the power spectrum changes in sterile neutrino models. Also, previous works have not explored the effect of a smaller field-of-view on the power-spectrum measurement.
We discuss the this approach and its limitations in the following paragraphs.

\subsection{Lensing convergence of subhaloes} \label{sec41}

\begin{figure*}
\includegraphics[width=1.\hsize]{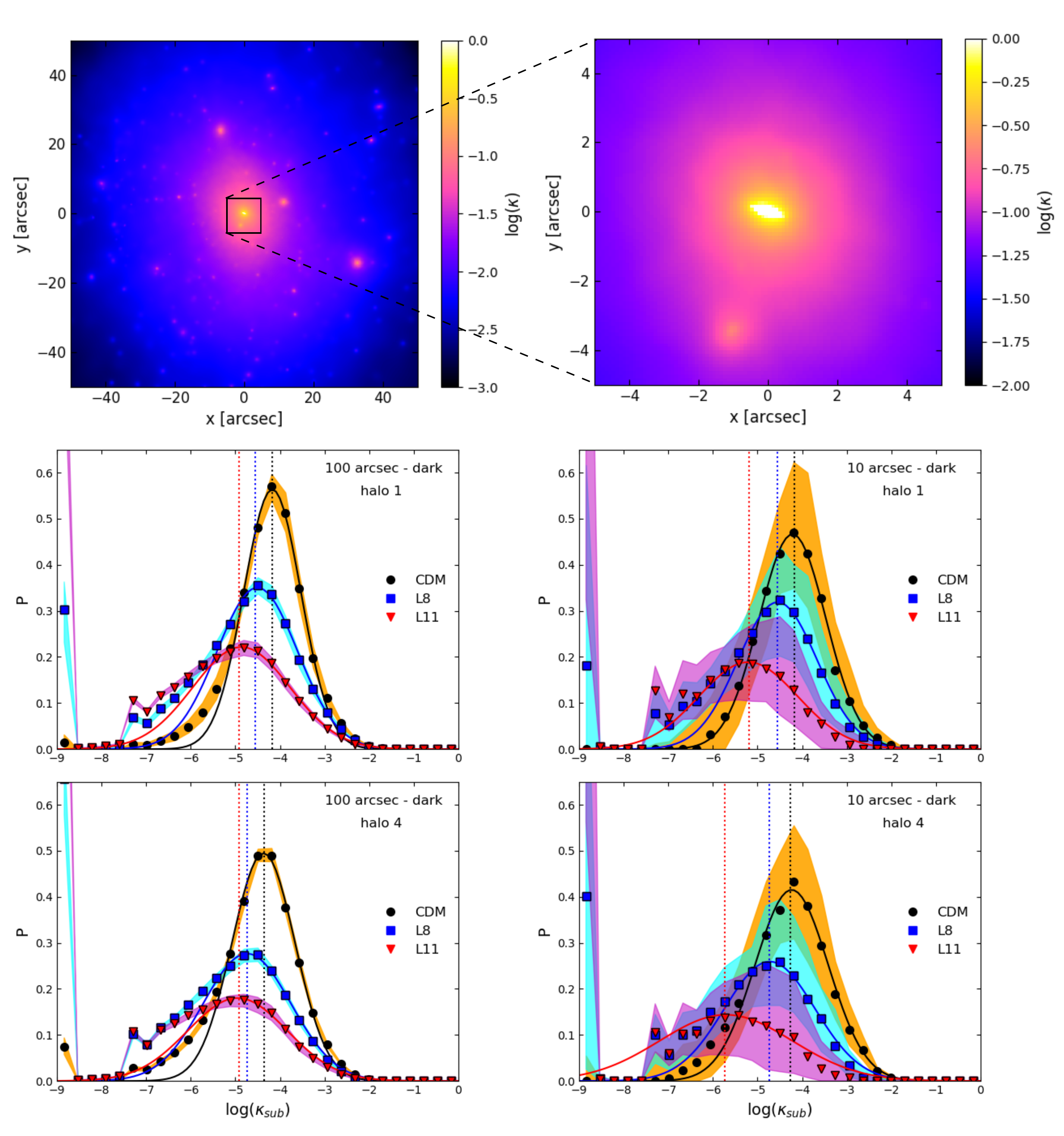}
\caption{\emph{Top row}: examples of the total convergence distribution in the two fields-of-view (100 in the left-hand panel and 10 arcsec on the right) for the CDM scenario. The colour-scale of the convergence is different in the two panels, in order to better describe the population in each case. \emph{Second and third row}: normalized distribution of $\kappa_\rmn{sub}$, at 100 (left) and 10 arcsec (right), for the most and least massive haloes in the sample (i.e. halo 1 and 4). Each distribution has been constructed with 200 random projections. The black, blue and red points stand for the median distribution of CDM, L8 and L11 scenarios, while the interquartile region (0.25-0.75) is shown by the coloured regions. The solid lines of corresponding colour show instead the best fit Gaussian distribution to each data-set; finally, the dashed vertical lines show the position of the peak. The pixel resolution is the same for all maps (0.01 arcsec) and different FoVs are obtained by cutting the 100 arcsec map. Since we do not include the background matter distribution of the halo, some pixels have a convergence equal to zero (especially in the L8 and L11 runs), which results in the last point of the histogram at $\kappa_\rmn{sub}\simeq 10^{-9}$. 
\label{conv}}\end{figure*}

The middle and bottom panels of Figure \ref{conv} show the logarithmic distribution of $\kappa_\rmn{sub}$, for the most and least massive haloes in our sample respectively. The points show the median distribution of each model, while the shaded regions enclose the 0.25 and 0.75 quantiles. 

In all of the panels, the solid lines show the best fit Gaussian distributions to the data, which reproduce the median well at the high-$\kappa$ end and around the peak, but fail at the low-$\kappa$ tail in all cases. This implies that the overall subhalo convergence distribution is well represented -- in particular in the CDM case -- by a Log-normal function in real space. 
By comparing the two panels, it is interesting to note that, while the variation between different projections becomes much larger in the 10 arcsec maps, the position of the peak  remains the same for the CDM and L8 models and changes only slightly for L11 (Table \ref{gauss_par}). The two haloes differ by almost one order of magnitude in mass (see Table \ref{tab_sim1}): the difference in the subhalo convergence distribution is however much smaller, indicating that the measure is quite stable with mass. In Table \ref{gauss_par}, we list the mean values for the mean and dispersion of the best-fit Gaussian through the whole sample for both sets of maps.

Since subhaloes are a diffuse component, and not as localized as the main galaxy convergence, it could still be possible to measure the true shape of their convergence distribution from small fields-of-view, assuming that the main lens convergence can be measured correctly. However, since the subhalo convergence is not observed directly, but reconstructed as a byproduct of the observational analysis by comparing the real data with the reconstructed lens model, it is also an approximation to assume that any small scale feature is caused by subhaloes \citep{bayer18,vegetti14,ritondale19b}.

\begin{table}
    \caption{Average parameters -- mean $\mu$ and dispersion $\sigma$ -- of the Gaussian best-fit to the substructure convergence distribution from the 100  and 10 arcsec maps (see Figure \ref{conv}) of the whole sample.
    \label{gauss_par}}
    \centering
    \begin{tabular}{c|c|c|c|c}
    \hline
    model & $\mu$ (100) & $\sigma$ (100) & $\mu$ (10) & $\sigma$ (10) \\
    \hline
    CDM & -4.27  & 0.67 & -4.22 & 0.78 \\
    L8 &  -4.62 & 0.94 & -4.63 & 0.95 \\
    L11 & -4.89 & 1.10 & -5.45 & 1.30 \\
    \hline
    \end{tabular}

\end{table}

\subsection{Convergence power spectrum}\label{sec42}

In general, the convergence power spectrum $P(\ell)$ of subhaloes is characterized by a number of features: first of all, the normalization depends on the overall convergence in subhaloes, which is by construction the main difference between the three models that we consider.  As already shown in \citet{brennan18}, the normalization of the subhalo power-spectrum is dominated by the highest mass subhaloes and thus removing them increases the relative difference between different models, as happens when excluding the `luminous' subhaloes from our calculation. As a test, we computed the power spectra using maps that contain all the subhaloes and we also find that in this case the median $P(\ell)$ is essentially identical across the three models. Moreover, since CDM and WDM models differ the most at the lower mass end of the subhalo mass function, it is a natural consequence that the relative difference increases when considering only low-mass subhaloes.

\begin{figure*}
\includegraphics[width=0.44\hsize]{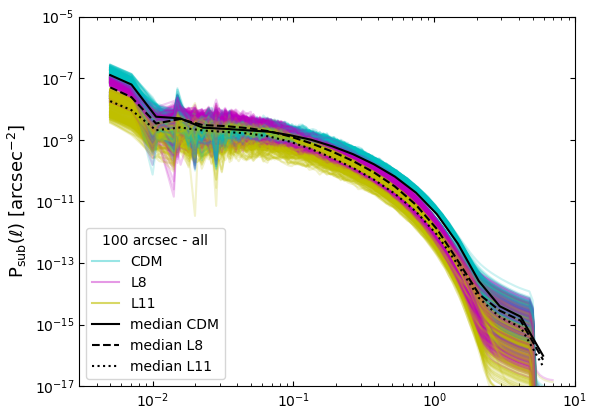}
\includegraphics[width=0.44\hsize]{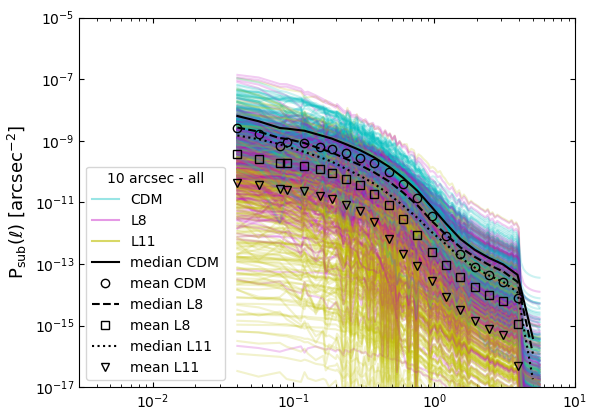}
\includegraphics[width=0.44\hsize]{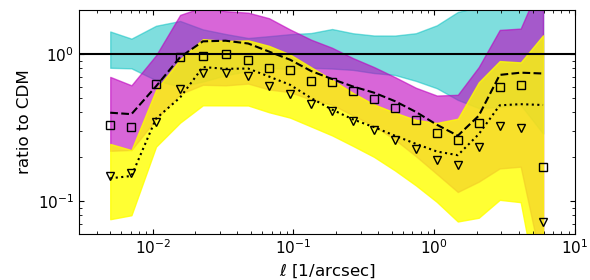}
\includegraphics[width=0.44\hsize]{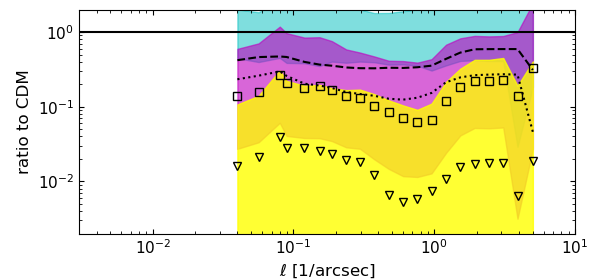}
\caption{Substructure power spectrum, measured from the 100 (left) and 10 (right) arcsec maps. Different projections (200 per halo in total) are shown in the top panels by the thin cyan (CDM), magenta (L8) and yellow (L11) lines, while the median $P(\ell)$ corresponds to the solid, dashed and dotted black lines. The two lower panels show the ratio of the median $P(\ell)$ of each model to the CDM one, and the coloured bands enclose the region between the 0.25 and 0.75 quantiles. We show both the median (dashed/dotted lines) and the mean (squares/triangles) values in both cases, obtained from the whole sample of haloes/projections.} \label{Pk_sub}
\end{figure*}

Figure \ref{Pk_sub} shows the power spectra measured within 100 (left) and 10 arcsec (right), calculated using a module from {\sc MOKA} \citep{giocoli12a}. Each thin coloured line shows the measurement for one projection/halo, while the solid, dashed and dotted lines show the median values. The lower panels show the ratio of each median power-spectrum to the CDM equivalent. The discriminating-power of the $P(\ell)$ seems to suffer from the limited field of view more than the convergence distribution does, as the measurements from different 10 arcsec maps scatter much more than those obtained from the 100 arcsec maps. 

We find that while the large scale modes ($\ell<0.03~\mathrm{arcsec}^{-1}$) are substantially identical for all the projections and for all three models, the power spectrum varies significantly at intermediate and small scales.  The mean power spectrum is lower for the sterile neutrino models with respect to CDM, showing a signature of the lack of small scale structures. We note that not only is the normalization different in the three scenarios, but also the slope of the power-spectrum for $k\geq 0.05$. The high-$k$ slope is determined by the density profile and thus the difference between the models is an indication of the different properties of the subhalo profiles. 

However, the scatter from different projections is larger than the difference between the medians and values from the different models overlap, as in previous works \citep{brennan18}. This suggests that it is virtually impossible to distinguish between the three models by analysing only a few lensing images. The scatter is smaller in the right-hand panel of Figure \ref{Pk_sub} than in the smaller field of view, but it is relevant in both cases.

Our findings suggest that, using the power spectrum, it might be possible to distinguish between CDM and very warm WDM models, but not the relatively cold sterile neutrino models presented in this work. As the differences increase at progressively lower masses, higher resolution simulations would be required to measure the power spectrum of subhaloes with $M\leq 10^{8}\rmn{M_{\odot}}$, which might be more significant. At the same time, we anticipate that future observations, that will allow us to detect those low masses -- and at the same time identify with more precision the most massive subhaloes, and subsequently to include them in the lens model -- will prove more important for power-spectrum studies than current data sets.

\section{Detection expectations} \label{sec_det}

As shown in \citet{despali18} and \citet{li16b}, subhaloes are only a fraction of the total number of low-mass haloes that contribute to the convergence in the strong lensing images: isolated field haloes, located along the line of sight, constitute at least 50 per cent of the total number of perturbers for any combination of source and lens redshift. Thus, they play a fundamental role in constraining the dark matter model and have to be taken into account. Here we provide approximate detection expectations through gravitational lensing including both field haloes and subhaloes. We assume that the halo mass function suppression in sterile neutrino models is the same as that of subhaloes, as in many previous works \citep{schneider12,lovell14,despali18,vegetti18}. We leave an explicit verification of this assumption to a follow-up paper. We  use the \citet{sheth99b} CDM halo  mass function parametrisation, with  the  best  fit  parameters appropriate  for  the  Planck  cosmology as calculated by \citet{despali16}. We then use the best fit parameters $\gamma$ and $\beta$, obtained by fitting Equation~\ref{massf_eq_2} to the ensemble of the sterile neutrino runs (see Figure~\ref{submf} and Table~\ref{tab_par}) and we use them to estimate the total expected number of perturbers, i.e. low-mass subhaloes and field haloes with masses $M_\rmn{min}\leq M\leq 10^{10}\rmn{M_{\odot}}$. $M_\rmn{min}$ is defined according to the lowest detectable mass from observational data, as detailed in the following paragraph.

We provide predictions for a representative sample of observed lenses, which reproduce HST data in terms of point-spread-function, resolution and signal-to-noise level appropriate for single-orbit observations - or artificially improved versions of the same data, as discussed below. We consider a sample of $\simeq30$ lenses, with lens and source redshifts in the range $0.2\leq z_\rmn{l}\leq 0.5$ and  $1\leq z_\rmn{s}\leq 2.5$.
This resembles the configurations of state-of-the art samples, such as the SLACS  \citep{bolton06,vegetti14,vegetti18} and BELLS GALLERY lenses \citep{shu16a,ritondale19a,ritondale19b}. 
The mean lowest detectable mass in the considered sample is $<M_\rmn{low}\simeq 5\times 10^{9}\rmn{M_{\odot}}$, while its minimum is $M_\rmn{low,min}\simeq 10^{8}\rmn{M_{\odot}}$ \citep[possible only in less than one per cent of the relevant area -- the distribution lowest detectable mass pixel-by-pixel for realistic data can be see in ][]{ritondale19b}.\footnote{We remind the reader that the data sensitivity in this case corresponds to the lowest detectable mass of (sub)haloes that can influence the surface brightness distribution of the lensed arcs. In particular, a detection is generally considered reliable if the lens model including a secondary clump is preferred to the smooth lens model with a difference in log evidence of $\Delta\log E\geq 50$ \citep{vegetti09,vegetti10,vegetti12,vegetti14}.  The exact value of the lowest detectable mass varies pixel-by-pixel, depending on the position of the secondary pertuber in the image plane.} 

\begin{figure}
    \centering
    \includegraphics[width=\hsize]{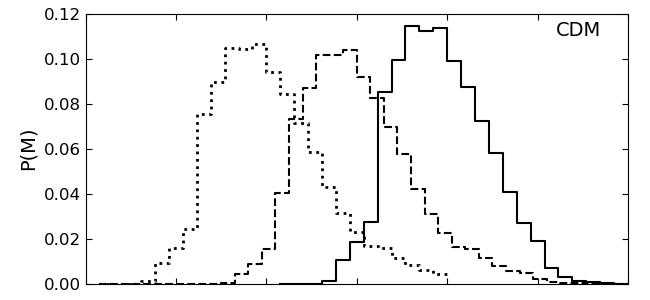}
    \includegraphics[width=\hsize]{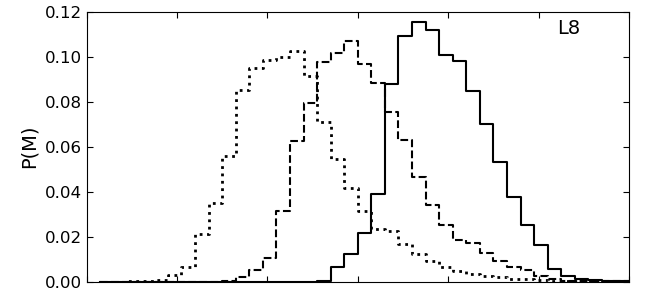}
    \includegraphics[width=\hsize]{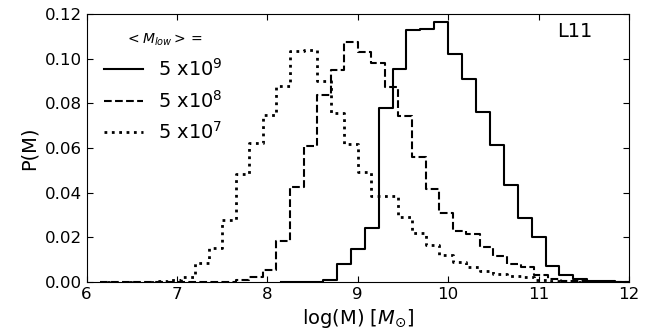}
    \caption{The probability of detecting a certain perturber mass $M$ in the joint sample of SLACS and BELLS lenses, for CDM (top), L8 (middle) and L11 (bottom panel), normalized by the peak probability for each $M_\rmn{low}$. The solid histograms show the distribution of detectable masses with the current data possibilities, while the dashed (dotted) lines are predictions for lowest detectable masses one (two) orders of magnitude lower in mass.}
    \label{pdet}
\end{figure}

In order to make some predictions and overcome the current data quality limitations, we look at the distribution of detectable subhaloes and field haloes for each sample, given three different mean values for the lowest detectable mass: the realistic one ($5\times 10^{9}\rmn{M_{\odot}}$) and two improved values, one and two orders of magnitude lower ($5\times 10^{8}\rmn{M_{\odot}}$ and $5\times 10^{7}\rmn{M_{\odot}}$ ). These would in practice correspond to HST data with a higher signal-to-noise or longer exposure times, given that the resolution is kept fixed. Figure \ref{pdet} shows what are the masses most likely to be detected in each model (different panels) and for these three cases (different lines). In the first panel, the CDM distribution moves almost self-similarly to lower masses when $M_\rmn{low}$ decreases, consistently with the constant slope in the halo and subhalo mass function. On the other hand, for sterile neutrino models (second and third panels), masses below $M\simeq10^{8}\rmn{M_{\odot}}$ are strongly suppressed and thus the peak remains above this value in all cases; moreover, the shape of the distribution for lower $M\simeq10^{8}\rmn{M_{\odot}}$ changes, with a longer and lower low-mass tail. At the same time, while the three solid histograms are practically identical, the dotted ones present clear differences: at this sensitivity, the three models could be distinguished. This is consistent with the results from 
\mbox{\citet{vegetti18}} and \mbox{\citet{ritondale19b}}, who found that with currently data it is not possible to put stringent constrains on the nature of dark matter: they can only safely exclude sterile neutrino models with $m_{s}<0.8$~$\rmn{keV}$, equivalent to $M_\rmn{hm}>10^{12}\rmn{M_{\odot}}$. 

Increasing the number of lenses -- at fixed data quality -- would result in a linear increase in the total number of detections, but would not alter the distributions in Figure \ref{pdet}. On the other hand an improvement in the lowest detectable mass -- i.e. moving to lower values of $M_\rmn{low}$ -- would allow us to reach the regime where the three models are different from each other, and thus distinguishable. We plan to extend these results, accurately modelling the effect of resolution and signal-to-noise, in a follow-up paper.

\section{Discussion and conclusions} \label{sec_discussion}

In this work, we used zoom-in simulations with cold or sterile neutrino dark matter to explore the properties of the subhalo population and to characterize the differences between these models in the framework of strong gravitational lensing. We considered two different sterile neutrino models, which could be related to the production of the 3.55~keV unexplained X-ray line observed in clusters of galaxies \citep{bulbul14}, M31 \citep{boyarsky14a}, the Galactic Centre \citep{boyarsky15} and the Milky Way halo outskirts \citep{boyarsky18,cappelluti18}. In both models, the particle mass is $m_{s}$=7 keV, but the lepton asymmetry has two different values -- $L_{6}=8,11.2$ -- which result in a different suppression of low-mass structures (see Figures \ref{images} and \ref{submf}). We resimulated four haloes hosting massive elliptical galaxies - i.e. typical lens galaxies at redshift $z\simeq 0.2 - 0.5$ - which have never studied in sterile neutrino hydrodynamical simulations. Here we summarise our findings and discuss the implications for gravitational lensing studies. 

Gravitational lensing, and in particular the gravitational imaging technique \citep{vegetti09} has the unique ability of detecting dark low-mass haloes, which are not visible as satellites of the main galaxy (or as low-mass galaxies along the line of sight). Through this technique it is possible to detect individual haloes \citep{vegetti10,vegetti12,hezaveh16}, but also to put constraints on the halo and subhalo mass functions \citep{vegetti14,vegetti18,ritondale19b}. For this reason, the results from simulations and theoretical model are of fundamental importance to interpret observational results.

We started by looking at the properties of the four main haloes in the three different models. We find that the main halo properties, such as total and stellar mass, virial and effective radius, are very similar in the three models. Moreover, the dark matter fraction only differs in the innermost part of the halo ($r\leq 0.05r_\rmn{vir}$), where it is 20 per cent lower in sterile neutrino models. However, with only four haloes, we do not have enough statistics to model this effect precisely. The most important difference remains the number of dark low-mass subhaloes, which are suppressed in sterile neutrino models, while the number of luminous satellites is similar (see Table \ref{tab_sim1}).

Then, we studied the properties of the substructure population, in terms of mass function (Sec. \ref{sec_mf}), spatial distribution (Sec. \ref{sec_rad} and \ref{sec_dmfrac}) and density profile (Sec. \ref{sec_prof}). Sterile neutrino models are more complex than other WDM candidates, due to the combined effect of $m_{s}$ and $L_{6}$ changing both the matter power spectrum cut-off wavenumber and slope in a non-trivial fashion. For this reason, a two parameter fit is needed to accurately reproduce the subhalo mass function. In particular, the best-fit expression from \citet{schneider12} underestimates the number of low-mass subhaloes in our simulations. We provide the best fit parameters for the L8 and L11 subhalo mass functions in Table \ref{tab_par}. The fact that the subhalo mass function parametrisation depends on the specific sterile neutrino model (as Equation \ref{massf_eq_2}) has an important implication: even measuring the subhalo mass function directly from observational data, as in  \citet{vegetti18}, might not be sufficient to constrain the sterile neutrino model correctly, given the degeneracy between $M_\rmn{hm}$ and $L_{6}$. However, some assumptions can be made to reduce the parameter space: for example, restricting the sterile neutrino mass to 7.1~keV, as we do here, would allow us to put constraints on $L_{6}$ for this particular model, or the other way around. In this perspective, a more precise parametrisation (such as the values in Table \ref{tab_par}) will help the comparison with observational results.

The normalization of the subhalo mass function scales with redshift and host halo mass, but the exact number of subhaloes can still vary from system to system. It has been shown that strong lensing studies that try to constrain the projected dark matter fraction in subhaloes $f_\rmn{sub}$ (within a certain radius) and the WDM half-mode mass $M_\rmn{hm}$ simultaneously \citep[e.g.][]{vegetti18,gilman19,hsueh19} find that these two quantities are correlated: similar results can be obtained by increasing $f_\rmn{sub}$ in a warmer model, or decreasing $f_\rmn{sub}$ in a colder dark matter model, as the two contributions balance. 
Previous observational \mbox{\citep{vegetti14,vegetti18}} and numerical \citep{xud15,despali17b} works have measured $f_\rmn{sub}$, finding consistent values for CDM of the order of $f_\rmn{sub}\simeq0.004-0.006$ for subhaloes of mass $10^{6}\leq M_\rmn{sub}\leq 10^{9}\rmn{M_{\odot}}$. The presence of baryonic physics suppresses the number of subhaloes, by different amounts depending on the details of the implementation, which can help to exclude unrealistic physical models \mbox{\citep{despali17b}}. Given that the same baryonic physics model has been used for all the runs, its effect is of the same order \citep{lovell17b}. In this work we calculated the projected number of subhaloes as a function of the distance from the centre, both for CDM and sterile neutrino models, as has been done previously for CDM alone \citep{xud15,despali17b}. These are flat (as a function of radius) in the inner $\simeq$30 kpc and lower in the sterile neutrino models, with a difference from CDM that increases at low subhalo masses, i.e. for $M\leq10^{8.5}\rmn{M_{\odot}}$. These results are broadly consistent with previous works based on WDM simulations, both in the form of sterile neutrinos \citep{lovell16} and thermal relics \citep{schneider12,lovell12,lovell14,ludlow16}. However, none of these has addressed specifically the host haloes of massive early-type galaxies in sterile neutrino hydrodynamical simulations, as we do in this work. 
We then measured the dark matter fraction in subhaloes as a function of radius, finding that it is lower towards the centre for the sterile neutrino models than in CDM, similarly to what was found by \citet{lovell14} for thermal relic WDM models. 

Finally, we measured the subhalo density profiles and calculated the best-fit Einasto profile. We found that the profiles are shallower towards the centre in sterile neutrino models, leading to central densities suppressed by 10 to 50~per~cent. For this reason, the subhalo concentrations are lower than in CDM by up to 30 per cent, for subhaloes with masses $M\geq 3\times10^{8}\rmn{M_{\odot}}$. 

In Section \ref{sec_lensing}, we looked at the differences in the lensing signal of subhaloes between the three models by analysing the distribution of subhalo lensing convergence and its power spectrum. We created convergence maps by ray-tracing through 200 random projections for each halo; for this, we used the particles belonging to non-luminous (and non-spurious) subhaloes. The subhalo population contributes to the total lensing convergence at different scales and in general the subhalo convergence is qualitatively well fitted by a Log-Normal distribution -- with departures from gaussianity at the edges -- in all models. In Table \ref{tab_par}, we list the best-fit parameters of the Gaussian distributions that fit the subhalo convergence in logarithmic space. These distributions describe the subhalo convergence well both when the measurement is done with the subhalo population of the entire halo (i.e. with convergence maps of $\simeq$100~arcsec on a side), and when considering only the inner part of the halo in projection ($\simeq$ 10~arcsec), closer to the location of the lensed images.

Similar information can be obtained from the power spectrum of subhalo convergence: the total power is lower in sterile neutrino models (Figure \ref{Pk_sub}), as well as the slope at $k\geq 0.1$~$\rmn{arcsec}^{-1}$: this is due to the different properties of the subhalo profile and to the different relative contribution of high and low mass subhaloes, since the former are present in similar numbers. One limitation of the power spectrum approach is that the variation between different projections is larger -- or at best of the same order of magnitude -- of the intrinsic differences between the power spectrum in the three models. However, the mean and median power spectra show a consistent difference, both when looking at the whole subhalo population and at a 10 arcsec field of view close to the halo centre. In this last scenario, the mean CDM power-spectrum remains consistent with the measurement done on the population within the virial radius, while the sterile neutrino models present a much larger variation and a lower mean -- simply due to the lack of structures on small scales in part of the projections. The power-spectrum is dominated by the high-mass subhaloes and thus removing them would increase the difference between the model. In order to measure the power due to the smallest structures alone, in the analysis of observational data one needs to identify the more massive subhaloes and explicitly include them in the model. Moreover, higher resolution observations are needed to probe the low-mass end of the mass function, where the models differ significantly from each other and to overcome the scattering due to different projections and a limited field-of-view. At the same time, the resolution of our simulations does not allow us to measure the power spectrum of subhaloes with mass $M\leq 10^{8}\rmn{M_{\odot}}$ reliably. 

Finally, in Section~\ref{sec_det} we follow the approach from \citet{despali18} and \citet{li16b} and provide detection expectations for a realistic sample of observed lenses, which resembles the configuration of the SLACS \citep{bolton06,vegetti14} and BELLS GALLERY \citep{shu16a,ritondale19a,ritondale19b} lenses. We consider both the current data-quality, in terms of resolution and signal-to-noise -- with an average detection limit around $M\simeq 5\times10^{9}\rmn{M_{\odot}}$ --  and artificially improved data which would allow to detect haloes and subhaloes with masses one or two order of magnitude lower. We assume that both the halo and subhalo mass function are suppressed similarly, following Equation \ref{massf_eq_2}; we used the best-fit parameters from Table \ref{tab_par} and calculate the number of effective perturbers (both subhaloes and isolated haloes along the line of sight) for the two samples, following the approach from \citet{despali18}. We calculate the distribution of detectable masses, comparing its peak and shape across the three scenarios. We find that, at the current data sensitivity, it is not possible to put significant constraints on the nature of dark matter or distinguish between CDM and the two sterile neutrino models considered here, consistentl with previous works \citep{vegetti18,ritondale19b}. However, if the data sensitivity increased by one or two order of magnitude in mass -- i.e. if we could detect (sub)haloes of mass $M\simeq 5\times10^{8}\rmn{M_{\odot}}$ or $M\simeq 5\times 10^{7}\rmn{M_{\odot}}$ -- the same samples would allow to distinguish between CDM and sterile neutrino models. 

In our simulations, we have used the same baryonic physics model in all the runs. In this way, we can focus solely on the effect of sterile neutrinos on the abundance of small-scale structures, since the effect of a fixed baryonic model on the abundance of subhaloes is very similar in the three cases \citep[as shown also in][]{lovell17b}. However, it is important to point out that there is a large amount of uncertainty in the effect of different baryonic physics models on the mass function of subhaloes \citep{despali17} and on their survival rates \citep{kelley19}. This causes a degeneracy between the effect of baryons and WDM, which would require a full set of hydro simulations that combines variations in both parameter spaces to be fully understood. Future observations with instruments such as E-ELT and VLBI  will have sufficient sensitivity to probe the halo mass function below $<10^{7}M_{\odot}$; in this regime the effect of different dark matter models is expected to play the dominant role in shaping the mass function, and thus the uncertainties in the baryonic physics modelling will no longer be an issue.

To conclude, in this work we presented the analysis of hydroynamical simulations with sterile neutrino dark matter. The chosen models -- in both cases sterile neutrino with a mass of $m_{s}$ = 7 keV -- are among the few alternative dark matter models with a possible observational motivation \citep[i.e. the unexplained 3.5 keV line][]{bulbul14,boyarsky15,cappelluti18}. They are, however, colder than most of the warm dark matter models explored so far in simulations \citep{schneider12,lovell12,lovell14,li16}. We provide detailed parametrisation of the differences between these sterile neutrino models and CDM. We conclude that the difference in the number of low-mass subhaloes leaves a clear signature on the distribution of lensing convergence and in the subhalo power spectrum. These are, however, not yet detectable with current samples of observed lenses. Future observations, more numerous and with a higher sensitivity (which could be provided by an improved signal-to-noise, longer observational times or a higher resolution) will provide a promising way of discriminating between CDM and sterile neutrino models. 

\section*{Acknowledgements}

We thank the EAGLE collaboration for allowing the use of the EAGLE model in these simulations and Joop Schaye for useful comments.
This work was carried out on the Dutch National e-Infrastructure with the support of SURF Cooperative. MRL is  supported  by  a  COFUND/Durham  Junior Research Fellowship under EU grant 609412, and also acknowledges
support by a Grant of Excellence from the Icelandic Research Fund (grant number
173929$-$051). SV has received funding from the European Research Council (ERC) under the European Union's Horizon 2020 research and innovation programme (grant agreement No 758853).  RAC is a Royal Society University Research Fellow. Support for BDO was provided through the NASA ATP grant NNX16AB31G. This work used the DiRAC@Durham facility managed by the Institute for Computational Cosmology on behalf of the STFC DiRAC HPC Facility (www.dirac.ac.uk). The equipment was funded by BEIS capital funding via STFC capital grants ST/K00042X/1, ST/P002293/1, ST/R002371/1 and ST/S002502/1, Durham University and STFC operations grant ST/R000832/1. DiRAC is part of the National e-Infrastructure. We thank Ben Metcalf for giving us access to the lensing code {\sc GLAMER} and Carlo Giocoli for providing the {\sc MOKA} libraries. This research made use of Astropy, a community-developed core Python package for Astronomy \citep{astropy13}, of the matplotlib 
\citep{matplotlib} and NumPy packages. 

\bibliographystyle{mnras}
\bibliography{mnras_template.bbl}

\label{lastpage}
\end{document}